\documentclass{article}  
\usepackage{amsmath}
\usepackage{todonotes}
\usepackage{bm}
\usepackage{cancel}
\usepackage{xcolor}   
\usepackage{arydshln}

\usepackage{ulem}
\definecolor{grey}{RGB}{100,100,100}

\begin{document}                  



\title{A minimal G$\bar{\textrm{o}}$-model for rebuilding whole genome structures from haploid single-cell Hi-C data}










\date{} 
\maketitle                        

\begin{center} 
\textsc{S. Wettermann$^{a}$\footnote{S.W. and M.B. have contributed equally\label{x}}, M. Brems$^{a}$\textsuperscript{\ref{x}}, J.T. Siebert $^{a}$, G.T. Vu $^{a}$, T.J. Stevens$^{b}$, P. Virnau$^{a}$}\\
\ \\
$^a$ Department of Physics, Johannes Gutenberg University Mainz, \\ Staudinger Weg  $9$, $55128$ Mainz, Germany\\
$^b$ MRC Laboratory of Molecular Biology, Francis Crick Avenue, Cambridge Biomedical 
	Campus, Cambridge CB$2$ $0$QH, UK.  \\
virnau@uni-mainz.de
\end{center}

\begin{abstract}
We present a minimal computational model, which allows very fast, on-the-fly construction of three dimensional haploid interphase genomes
from single-cell Hi-C contact maps using the HOOMD-blue molecular dynamics package on graphics processing units. Chromosomes are represented by 
a string of connected beads, each of which corresponds to 100,000 base pairs, and contacts are mediated via a structure-based harmonic 
potential. We suggest and test two minimization protocols which consistently fold into conformationally similar low energy states. The latter are similar to previously published structures but are calculated in a fraction of the time. We find 
evidence that mere fulfillment of contact maps is insufficient to create experimentally relevant structures. Particularly, an excluded volume 
term is required in our model to induce the formation of chromosome territories. We also observe empirically that contact maps do not capture the chirality of 
the underlying structures. Depending on starting configurations and protocol details, one of two mirror images emerges. 
Finally, we analyze the occurrence of knots in a particular 
chromosome. The same knot appears in (almost) all structures irrespective of minimization protocols or even details of underlying potentials providing further 
evidence for the existence of knots in interphase chromatin.
\end{abstract}

     
\section{Introduction}
\label{sec:intro}


Knowledge of the three-dimensional structure of interphase chromatin is crucial for the understanding of gene regulation \cite{CremerNRG:2001} and other important processes such as
cellular differentiation \cite{DixonNature:2015} and the cell cycle \cite{NaumovaScience:2013}. At the large scale this information was first obtained from fluorescence in-situ hybridization (FISH) experiments showing that
interphase chromosomes are usually compact and form territories within the nucleus \cite{CremerNRG:2001,CremerCSH:1993,BrancoPLoS:2006}. The detailed structure, however, has only 
become accessible recently with the rise of chromosome conformation capture techniques such as 3C \cite{DekkerScience:2009} and derived methods like Hi-C \cite{LiebermannScience2009}. 
The latter applies high-throughput sequencing techniques to determine (non-sequential) spatial contacts between DNA strands from proximity ligation of DNA by sequencing ends and mapping them back onto a 
reference genome. With the contact map of single-cell Hi-C \cite{NaganoNature:2013,StevensNature:2017,NaganoNature:2017,FlyamerNature:2017,RamaniNM:2017,TanScience:2018}, the folded structure of two meters of DNA 
into the confines of a nucleus a few microns wide was revealed at scales down to 20,000 base pairs \cite{TanScience:2018}.

So far, there is still no physical model which is generally agreed upon as the currently accessible resolution is still above the nucleosome size and persistence
length. Various methods have been discussed recently to reconstruct structures from single cell Hi-C maps such as approaches based on 
Bayesian inference \cite{Rosenthal:2018}, manifold based optimization \cite{PaulsenPLoS:2015} or the minimization of 
polymer models \cite{NaganoNature:2013,StevensNature:2017,NaganoNature:2017,CarstensPLoS:2016,ZhangPNAS:2015,ZhangPRL:2016,ZhangBiophys:2017}. Note that numerous similar approaches have also been used on bulk Hi-C data 
i.e., contact maps derived from multiple cells, working under the assumption that contacts and characteristic features of structures appear simultaneously in all or at least many cells
(see, e.g., \cite{StefanoPLoS:2013,StefanoSR:2016,PierroPNAS:2016} and references therein.) 
Our single cell ansatz is based on a polymer model and aims at fulfilling two specific objectives. Firstly, we would like to construct a minimal model
with as simple a parametrization as possible. Secondly, we want to optimize the minimization procedure and place it on a modern computing platform using graphics
processing units (GPUs). The implementation of these two goals allows us to rebuild whole genome structures (as opposed to single chromosomes) with a resolution of 
100,000 base pairs almost instantaneously. 

In the following sections we describe our choice of potentials, simulation protocols and analysis tools to obtain the three-dimensional structure of haploid
mouse interphase chromatin from the contact map. We find that structures generated by our procedure are rather similar and belong to either one of two chiralities 
depending on starting conditions and protocol details. This is an indication that a given contact map in conjunction with our choice of potentials 
leads to a ground-state like conformation with a two-fold chirality. We find that the inclusion of an excluded volume term in our model is 
essential for the emergence of chromosome territories and plays a major role in the amount of self-entanglements observed. By choosing a rather large 
excluded volume term we are able to reduce but not prevent knots from occurring in the final structures.

\section{Model and simulation protocols}
\label{sec:model}

The task of obtaining a three-dimensional structure from a single-cell contact map is essentially an optimization problem which we approach by applying various concepts from computational statistical physics.
Chromosomes are modeled as chains of connected monomers, each corresponding to $100,000$ base pairs. This resolution was chosen to have enough beads with one or more contacts and ensure that in turn most of the 
structure is well-defined. The haploid mouse genome is therefore represented by $20$ chains of lengths ranging from $N\approx500$ to 
$N\approx2000$ monomers.
A general assumption of these models is that the chromatin path at a given scale can be approximated by a chain of uniform spherical monomers. Naturally, the smaller-scale conformation of the chromatin backbone is highly 
convoluted and there will be a vast distribution of underlying non-spherical conformations across the model. Nonetheless, the uniform, spherical bead approximation is a reasonable average: given it is isotropic, can 
accommodate a degree of uncertainty using harmonic distance restraints and because interphase chromosomes are suspected to have fractal-like chain crumpling. The latter is indicated by Hi-C contact probabilities 
\cite{LiebermannScience2009} and suggests there is some degree of self-similarity over a wide range of length scales with an overall compaction greater than a more extended, equilibrated polymer.
Beads which are in contact according to the experimentally determined contact pairs are also connected by harmonic springs which drives the system towards a native state. 
This concept is borrowed from protein folding simulations of so-called G$\bar{\textrm{o}}$-models \cite{Taketomi:1975, Socci:1995, Boelinger:2010, JarmolinskaJMB:2019}.\\ In $1975$, G$\bar{\textrm{o}}$ and coworkers suggested a two-dimensional
lattice model which captures aspects of protein folding. They defined $"$bonds$"$ between monomers which were to end up in spatial vicinity in the native state in an attempt to study dynamics of folding with Monte Carlo simulations.
Nowadays, the term is typically used for a class of generic polymer models in which $"$native bonds$"$ are defined between amino acids which are known to be in $"$contact$"$, i.e. close spatial proximity according to structural
information from the Protein Data Bank \cite{ProteinDataBank:2000}.
However, while the latter aim at revealing the folding trajectory towards a previously known native state we apply this idea to chromatin for which contact data is available to reveal the unknown $"$native$"$ state.
The actual trajectory is of no physical relevance in this case.
For our particular application, using Hi-C data from single cells, it is further assumed that the contacts can be satisfied by having spherical beads effectively touch, but not superimpose. This encompasses the uncertainty in the underlying chain path, as there can be many ways that a DNA ligation can be accommodated
between two beads, and leads to solutions of fairly uniform density (that may be compared with microscopy \cite{StevensNature:2017}) where the bulk of the spherical objects do not mix, even where they are known to have short-range contacts. 
\\This ansatz is by no means original and is a central element of almost all polymer-based minimization methods for the determination of chromosome structures because the contact potential is essentially responsible to enforce contacts.
Note that with this approach we strive to obtain a global structure based on local information. Non-local information on distances between various base pairs would likely improve the integrity of the global structure and could be
implemented naturally in this class of models. 
Our simulation protocols allow for a maximal exploration of conformations. The gradual increase of excluded volume from a
random starting configuration in protocol $1$, for example is inspired by equilibration techniques for polymer melts \cite{Auhl:2003}. Our approaches are also comparable to methods such as simulated annealing \cite{Kirkpatrick671} which also aim at locating low energy 
conformations or ground states. Potentials are chosen to facilitate this optimization task and should not necessarily be regarded as realistic coarse-grained representations of chromatin. 
To this end we define rather stiff bonds between adjacent beads and beads in contact which ensure that contacts and bonds are enforced. In addition, a rather large penalty is enforced in our excluded volume 
term for overlaps, which pushes beads without contacts away from each other but still allows for bond crossings in the expansion phase.

\begin{figure}[h]
    \centering
        \includegraphics[width=0.5\textwidth]{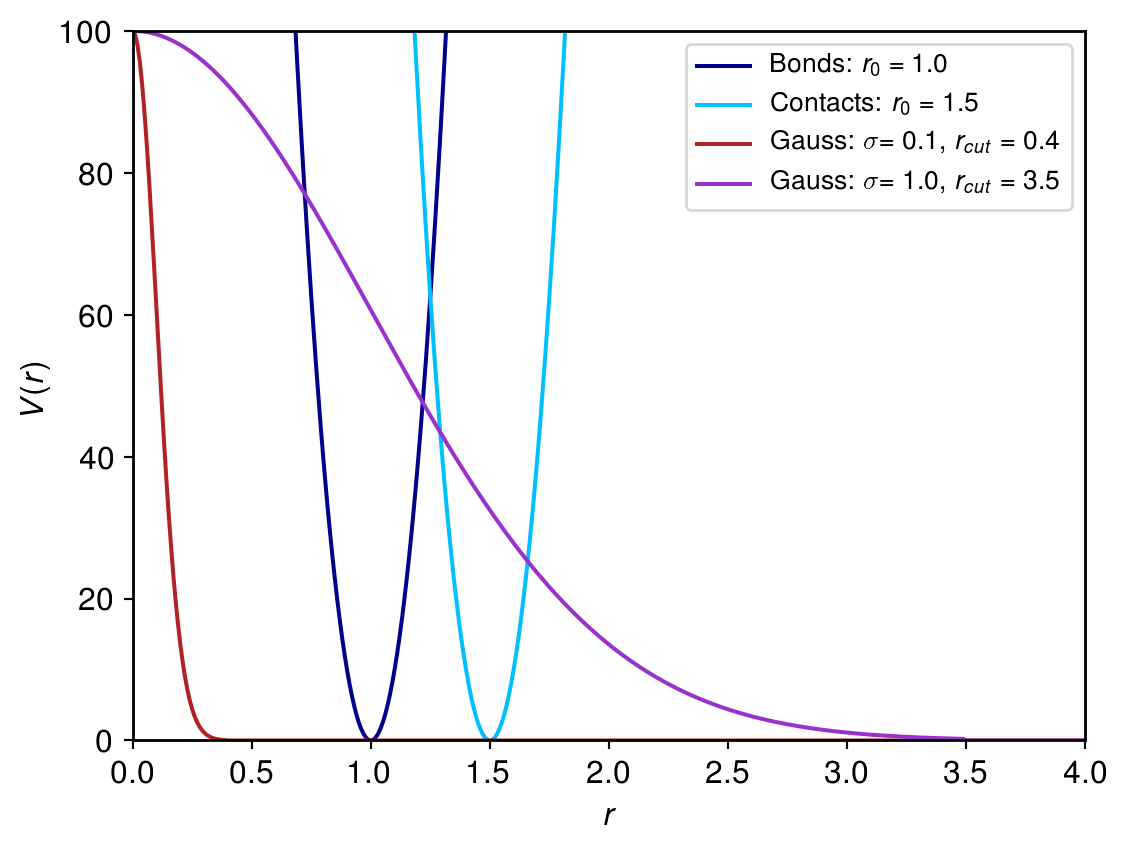}
	\caption{Bond, contact and (non-bonded) excluded volume potentials used for modeling chromatin.}
        \label{fig:potentials}
\end{figure}

Specifically, we use a generic bead-spring polymer model \cite{KremerGrest:1990} in which adjacent monomers are connected by a harmonic potential (Fig.~\ref{fig:potentials}):
\begin{equation}
\label{eq:harmonic} V(r) = \frac{1}{2} \kappa (r-r_{0, B/C})^{2},
\end{equation} 
where $\kappa=2000$ is the force constant and $r_{0,B}=1$ the preferred bond length. 
In all cases we refer to standard simulation units and do not mention them explicitly.
Contacts between beads are also enforced via a "native" harmonic potential with $\kappa=2000$.
The preferred distance, however, is somewhat larger ($r_{0,C}=1.5$) which increases 
similarities of final structures in terms of the root mean squared deviation (RMSD).  
Contacts are obtained from Hi-C data available at \cite{StevensNature:2017} and binned to sizes of $100,000$ 
base pairs. For the remainder of the paper we will focus on cell $2$ of this data set. From the binned data self-contacts and contacts between adjacent monomers are removed.
Multiple contacts between two beads are also not considered as the contact  potential of our 
minimal model does not depend on the number of contacts.
This reduces the number of contacts in cell $2$ of the data set from $79,569$ to $32,243$.

A Gaussian pair potential is acting between all particles that are neither adjacent nor in contact and pushes non-bonded monomers away from each other:
\begin{equation}\label{eq:gauss}
  V_{\textrm{\scriptsize Gauss}}(r) =
  \begin{cases}
  \epsilon \exp \left[-\frac{1}{2} \left(\frac{r}{\sigma}\right)^{2}\right] & r < r_{cut}  \\
  0 & r \geq r_{cut} 
  \end{cases}
\end{equation}
where $\epsilon=100$, whereas $\sigma=1$ and cut-off radius $r_{cut}=3.5$ (Fig.~\ref{fig:potentials}).
In the intermediate step of our simulation protocol the excluded volume is reduced ($\sigma=0.1$ and $r_{cut}=0.4$).  

It should be noted that energies in the model do not represent physical energies arising in chromatin. The sole purpose of these models and protocols is to find a ground-state-like conformation (similar to the native state of a protein)
which fulfills all (or almost all) contacts. In this sense the whole method can be considered as a kind of black-box algorithm to generate a structure which is compatible with the contact matrix of a given cell. Therefore, the resulting
structure is as far away from equilibrium as the original structure in the cell on which it is based upon. Furthermore, reference \cite{LiebermannScience2009} shows that the contact probability power law of the genome structures (which are very close to ours), 
and which also match bulk Hi-C, is inconsistent with a fully equilibrated polymer.  \\Constants in our potentials are selected to facilitate this purpose. Bond and contact potentials are harmonic which is typical
for generic bead-spring polymers. Prefactors are chosen to be extraordinarily large to strongly enforce contacts and bonds, but the exact choice of $\kappa=2000$ is somewhat arbitrary. The bond length of $r_{0,B}=1$
of our potential is a standard choice, again, while we opted to increase the optimal contact distance $r_{0,C}$ slightly to $1.5$ to introduce a second length-scale, which we find lowers the RMSD of the resulting structures.
Another motivation for this particular choice of $r_{0, B/C}$ was to enable comparisons with the model of \cite{StevensNature:2017}, which features similar optimal distances.\\ Below we will define two variants of our simulation protocol.
In contrast to more sophisticated methods \cite{Diao:2015} which aim at generating random walks in a confined volume, in protocol $1$ we simply place $25,719$ particles in a unit cube using a random, uniform distribution. 
To ensure maximum randomness particles are connected randomly to form $20$ linear chains corresponding to $20$ chromosomes with lengths ranging from $582$ and $1925$ particles.
The number of particles was not derived directly from the contact pairs, but obtained from \cite{StevensNature:2017} to enable a comparison with the reference structures. (Trajectories can be obtained upon request.)
From the starting configuration we proceed through the following steps:

\begin{itemize}
	\item $80,000$ time steps with bond and contact potentials but $\it{no}$ excluded volume
	\item $50,000$ time steps with bond, contact and $\it{reduced}$ excluded volume interactions
	\item $50,000$ time steps with bond, contact and $\it{full}$ excluded volume interactions
\end{itemize}

\begin{figure}[h]
	\centering
	\includegraphics[width=1.0\textwidth]{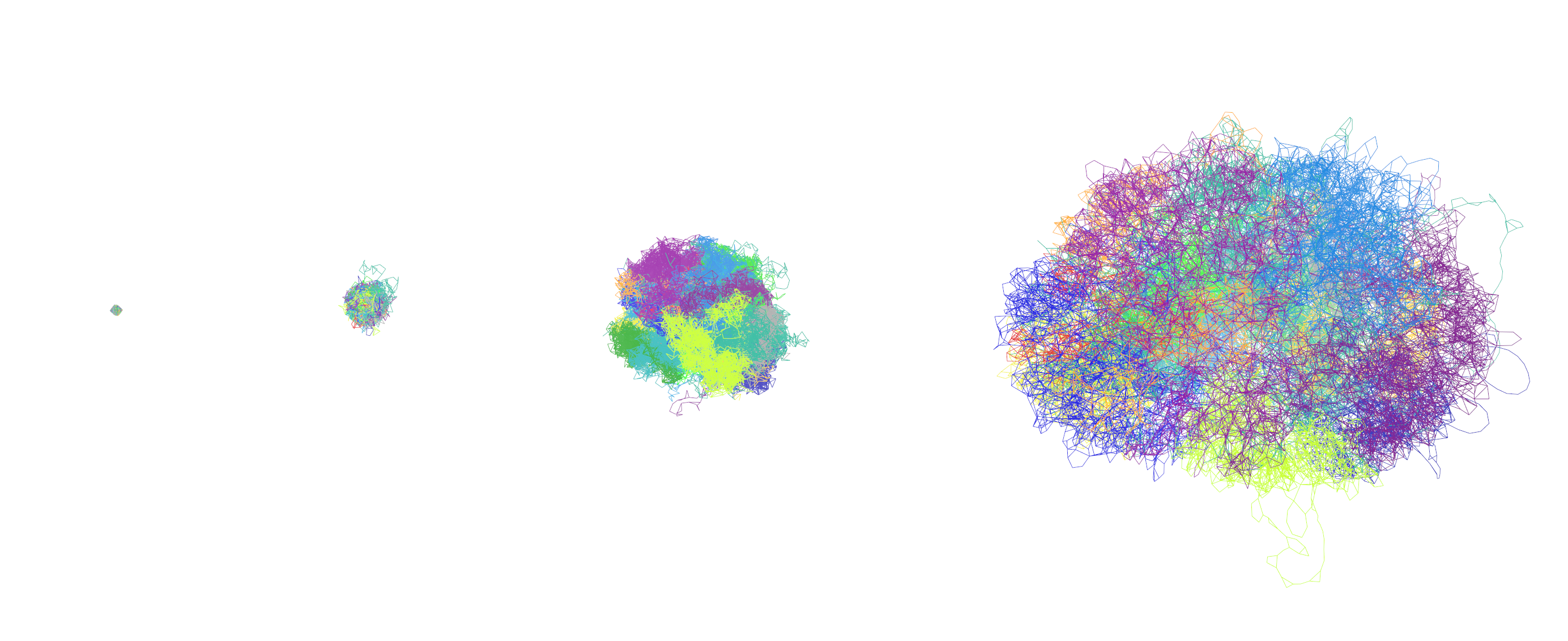}
	\caption{Starting configuration and snapshots taken at the end of each step in the first cycle of our simulation protocol: 
	a$)$ Particles are placed and connected randomly in a unit cube. b$)$ Bonds and contacts are enforced as described in the first step. c$)$ In the second step 
	the configuration expands  by introducing a rather short-ranged Gaussian excluded volume potential ($\sigma=0.1$ and $r_{cut}=0.4$). d$)$ 
	The full excluded volume ($\sigma=1$ and $r_{cut}=3.5$) is turned on in the third step. Different colors represent the individual chromosomes. 
	Figures were generated with VMD \cite{VMDPaper:1996}.}
	\label{fig:protocol}
\end{figure}

The number of time steps were chosen such that the structure can easily reach an energy plateau in each simulation step. Note that this number may need to be adjusted if different resolutions are considered or the underlying contact data is changed.
The cycle is iterated over and over again. The last frame of each cycle is saved and constitutes a new and independent configuration. The first or the first couple of frames are discarded because it typically 
takes two runs for the energy to converge. 
Note that having no or reduced repulsion in the early stages of the calculations has also been fairly common in simulated annealing of protein structures with distance restraints (e.g. from NMR \cite{NilgesNMR:1988}).
The expansion with reduced excluded volume in the second step was initially introduced to limit the amount of GPU memory used in conjunction 
with neighbor lists. Fig.~\ref{fig:protocol} shows the starting configuration and snapshots taken at the end of each step in the first cycle. Different colors indicate different chromosomes. 

For comparison, we also test a second protocol. Here we do not de- or inflate our configuration but turn on and off the contacts instead. For this protocol we start with a random walk of $25,719$ beads which is cut into $20$ chromosomes.
The actual cycle only consists of two steps:
\begin{itemize}
 \item $100,000$ time steps with bond and excluded volume interactions but no contact potentials
 \item $100,000$ time steps with all interactions
\end{itemize}

All simulations use HOOMD-blue \cite{Anderson:2008, Glotzer:2015} with a Langevin thermostat ($\gamma=1$, $k_{B}T=1$) and a time step of 0.001. We choose the implementation of the LBVH tree \cite{LBVH:2016} as our neighbor list which scales 
with particle number as opposed to the system volume. 

One cycle of protocol $1$ only takes about 100 seconds on a desktop PC with a NVIDIA 1050 graphics card and allows us to generate three dimensional structures from Hi-C contact matrices almost at once.
Note that this time includes a setup time for each step of the cycle which adds up to about $20-30$s. The time for the actual integration can of course be reduced further with a more powerful GPU. (With a RTX 2070 on our cluster 
one cycle only takes about $30$s).
Even though this is not a typical application of HOOMD-blue which thrives on long runs with hundred of thousands of particles on high-end GPUs, in this case it enables on-the-fly model and code development on an inexpensive desktop PC.


\section{Analysis of resulting structures}

Before analyzing structures in more detail we would like to discuss apparent features arising from Fig.~\ref{fig:protocol}. In the first step of the cycle (Fig.~\ref{fig:protocol}b) bonds 
and contacts are enforced, but monomers do not have any volume and can freely pass through each other with no costs involved. While such a configuration fulfills all contacts by construction it does
not exhibit chromosome territories. In a sense a globular intertwined configuration such as this is already a trivial (non-unique) solution to our optimization problem without the apparent features observed 
in experiments. This not only indicates that fulfillment of the contact map is insufficient to obtain experimentally relevant structures, it also suggests that we need to reformulate our task 
to account for beads that are not in contact. In our case this additional constraint is effectively enforced by the introduction of excluded volume which pushes beads away from each other if they
are neither adjacent nor in contact. As can be seen in Fig.~\ref{fig:protocol}c, adding a rather short-ranged excluded volume interaction already leads to the emergence
of chromosome territories which become more pronounced when full excluded volume interactions are enforced (Fig.~\ref{fig:protocol}d). Note that structures obtained as a solution of our optimization 
problem thus depend to some degree on the choice of our model parameters. As bead sizes increase the number of short-ranged contacts (not accounted for in the contact map) decreases naturally.
Enforcement of non-contacts via excluded volume interactions is, unfortunately, not entirely unproblematic as Hi-C contact pairs 
typically contain only a small fraction of potential contacts that actually exist in the real cell. These missing contacts are, at least in a good data set, often a consequence of sequence mapping problems with
repetitive genomic DNA and especially apparent at the centromers. The outward reaching arms which emerge in the final structure are also mostly a result of missing contacts.

In the following we will focus on the discussion of cell 2 of the data set provided by \cite{StevensNature:2017}. For simplicity, we will use the first simulation protocol to generate structures, i.e., we will initialize
the system once and run the cycle over and over again. For cell 2 (and most other cells in the data set) this approach preserves the chirality of structures across cycles and simplifies
our analysis. Chirality, which typically arises using protocol $2$ will be discussed in the next section.

\begin{figure}[h]
	\centering
	\includegraphics[width=0.46\textwidth]{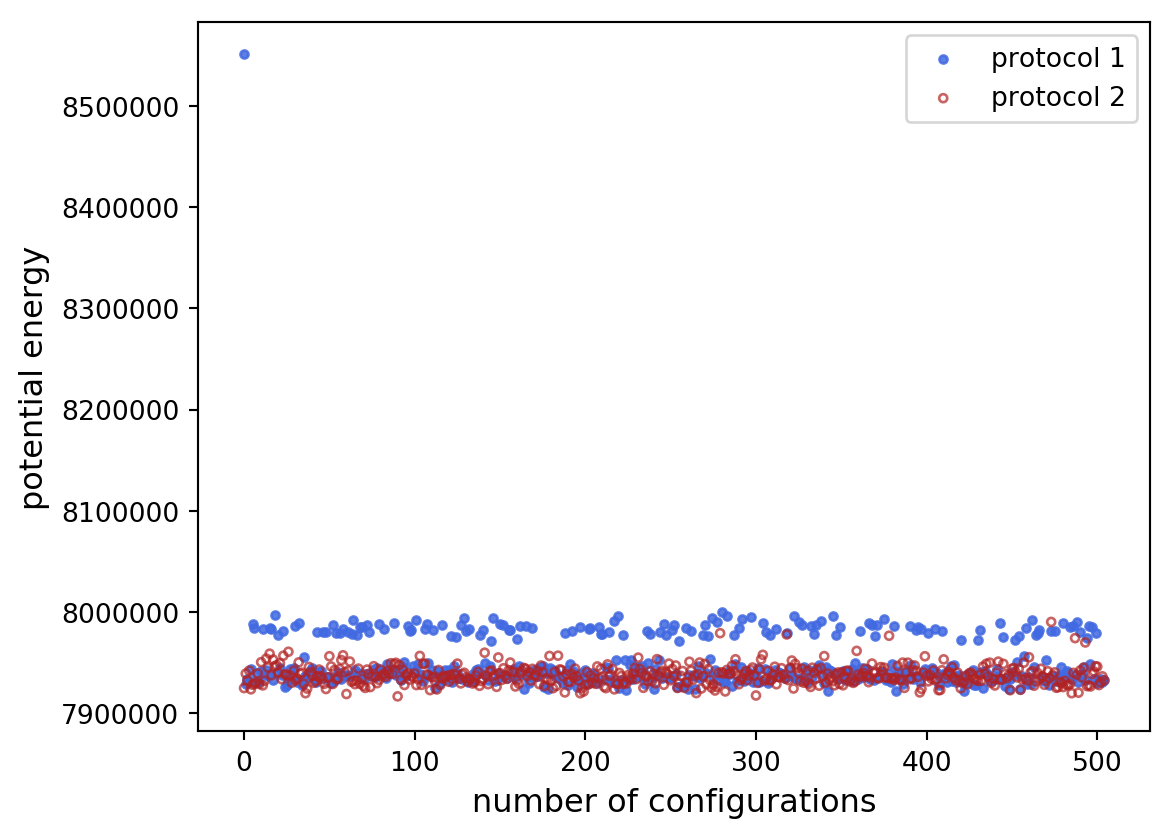}
	\includegraphics[width=0.445\textwidth]{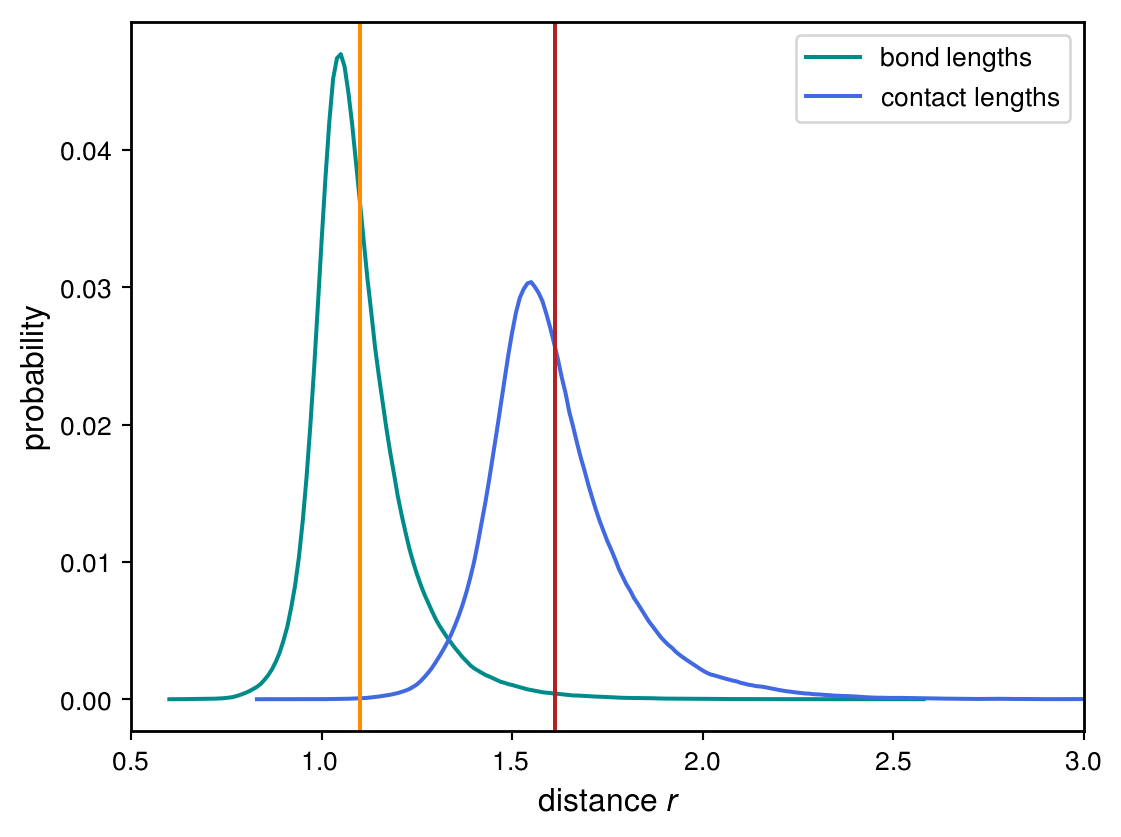}
	\caption{a$)$ Potential energy of the configuration corresponding to the last frame of each cycle. The energy quickly converges after the first cycles and fluctuates between two conformationally similar states.
	 b$)$ Distributions of bond and contact lengths. }
	\label{fig:energy}
\end{figure}

Fig.~\ref{fig:energy}a shows the potential energy of the final structure at the end of each cycle as a function of the number of cycles. At the end of the first cycle the energy is still somewhat larger which is
why this frame should be discarded before the analysis.\footnote{In fact, we will typically exclude the first five frames to be on the safe side.} In subsequent cycles the energy of the last frame of the cycle has converged and jumps between two conformations which only differ slightly.
Fig.~\ref{fig:energy}b shows the distribution of bonds and contacts. Both peaks and averages are shifted a little to the right of the potential minimum at $r_{0,B/C}$ indicating that some of the bonds and contacts
are overstretched. (This might be expected when approximating chromatin with spherical monomers.) Nevertheless, our step potentials manage to enforce constraints and the fraction of bonds and contacts which are overstretched beyond $2\cdot r_{0,B/C}$ is below $3\cdot 10^{-4}$ and $10^{-4}$,
respectively, which is significantly lower than reference simulations of cell $2$ in \cite{StevensNature:2017}. As our algorithm essentially enforces all contacts and bonds, all experimental contacts are also present in 
a matrix based on the generated structure. However, as experimental contacts only amount to a small fraction of actual contacts in the cell, we expect to find more contacts in our model structure.

\begin{figure}[h]
	\centering
	\includegraphics[width=0.5\textwidth]{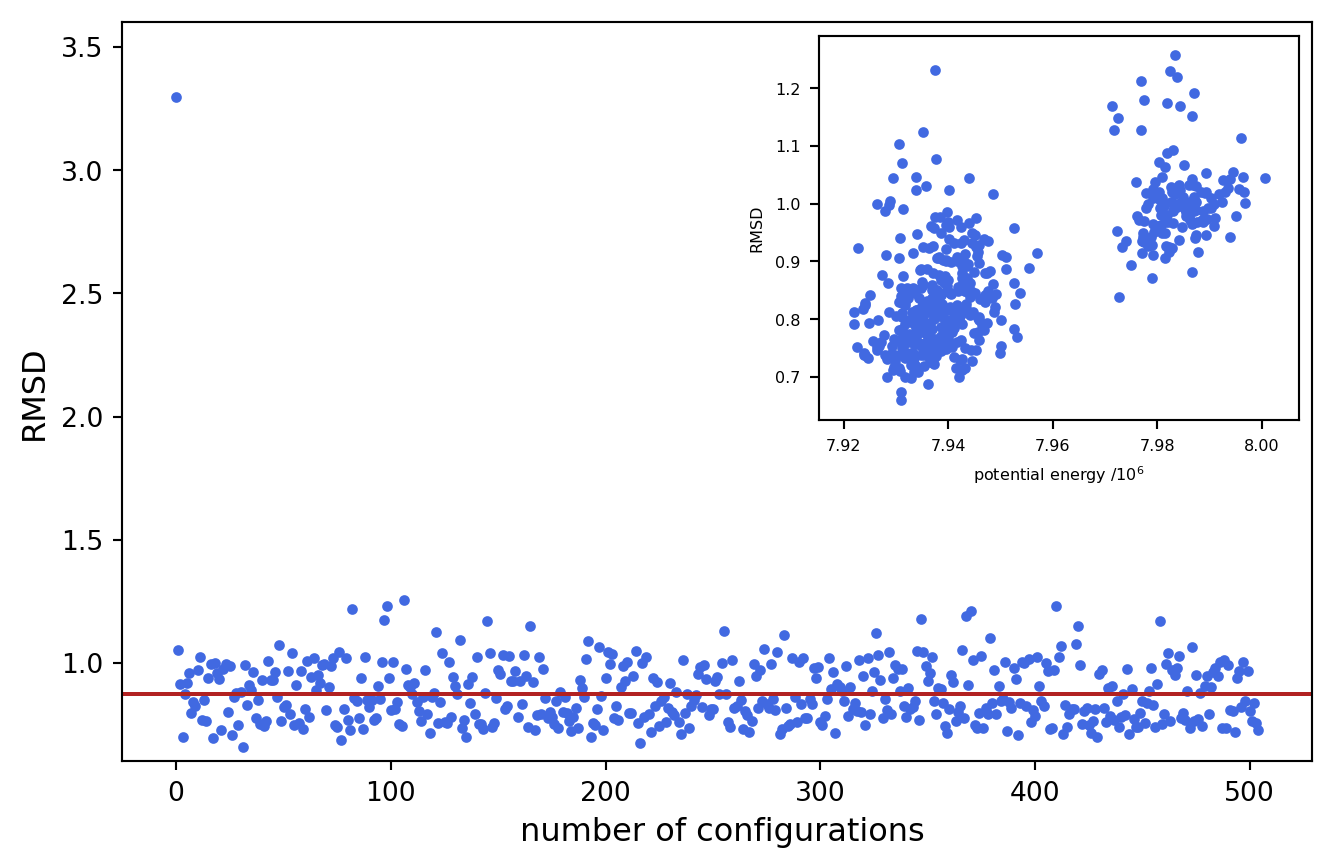}
	\caption{Root mean squared deviation from a pre-averaged structure for the final configuration of each frame (protocol 1).}
	\label{fig:contact_rmsd}
\end{figure}

In Fig.~\ref{fig:contact_rmsd} we analyze the similarity of structures by measuring deviations from an average structure.
To this end we determine the RMSD with respect to a pre-averaged reference structure using the RMSD trajectory tool of VMD \cite{VMDPaper:1996}:
\begin{equation} \label{eq:rmsd}
	\textrm{RMSD} = \sqrt{\frac{\sum_{i=1}^{N_\textrm{atoms}} (\vec{r_{i}}(t_{j})-\vec{r_{i}}(t_{0}))^{2}}{N_{\textrm{atoms}}}},
\end{equation} 
where $N_{\textrm{atoms}}$ is the number of monomers and $\vec{r_{i}}(t_{j})$ are the positions of the $i^{\textrm{th}}$ monomer of frame j and $\vec{r_{i}}(t_{0})$ the corresponding position in the pre-averaged reference frame.

As illustrated by the low average RMSD of 0.88 all structures are remarkably similar demonstrating that our method is indeed converging to very similar ground-state like conformations after the first cycle has been discarded.
The inset of Fig.~\ref{fig:contact_rmsd} reveals that the two competing final states are also similar in terms of RMSD. 
This result can be compared to reference simulations in \cite{StevensNature:2017}, which obtain a slightly smaller RMSD of 0.67, however not without pre-selecting structures.
(The hierarchical annealing protocol of the latter would discard outlier structures at an intermediate resolution/bin size which was required for the most sparse data sets when 
chromosomes sometimes adopted a chirality opposite to the rest of the structure and got stuck.)
In general, we recommend to regard resulting configurations as an ensemble of optimization runs and select only the states with the lowest energies. 
If, for example, we only consider the ten percent of all configurations with the lowest energies the RMSD drops to 0.76 for protocol 1.

\section{Chirality}

\begin{figure}[h]
	\centering
	\includegraphics[width=0.46\textwidth]{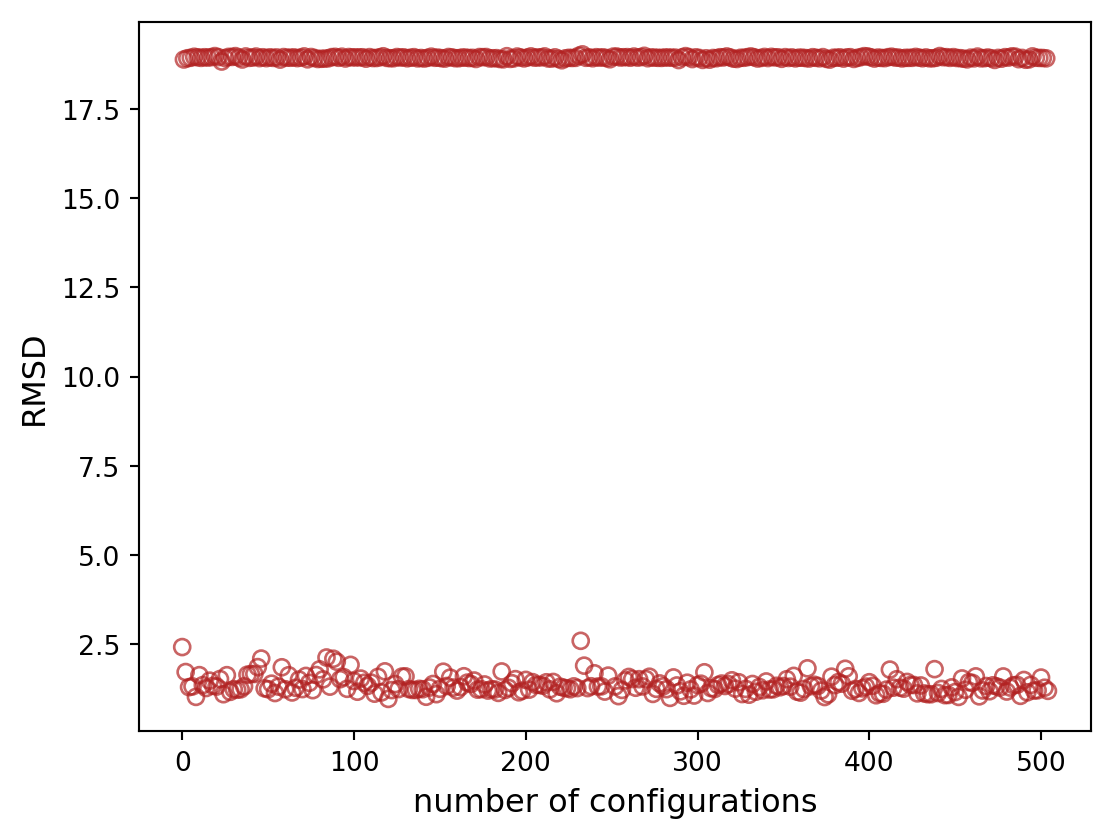}
        \includegraphics[width=0.445\textwidth]{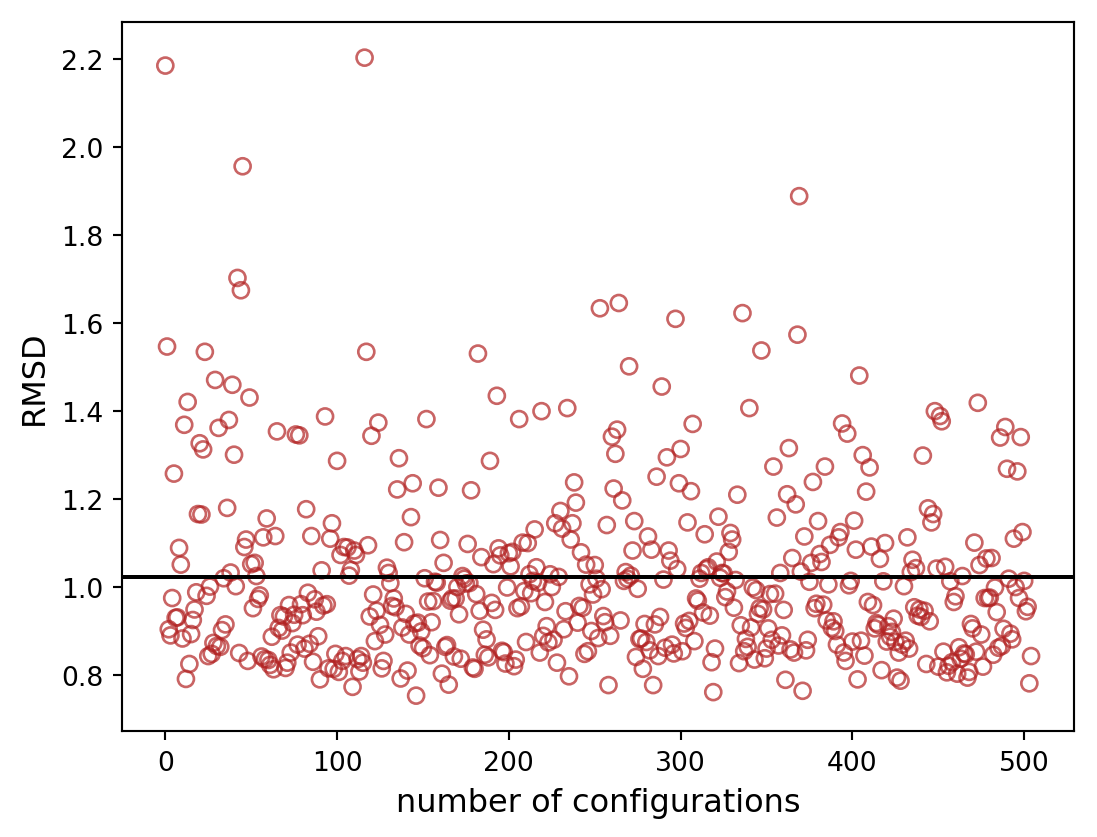}
	\caption{a$)$ RMSD arising from protocol 2 with respect to a specific configuration. b$)$ RMSD after mirroring transformation.}
	\label{fig:chiral}
\end{figure}

In this section we will discuss chiral structures which arise naturally if protocol 2 is applied or if we compare structures resulting from different starting configurations using protocol 1. 
To detect chiral structures we compute the RMSD with respect to a specific, but randomly chosen configuration. As shown in Fig.~\ref{fig:chiral}a, structures cluster into two groups with small and large RMSD,
respectively. If one multiplies, for example, all x-coordinates of one group by $-1$, both groups can be mapped onto each other (Fig.~\ref{fig:chiral}b) indicating that they are in fact mirror images.
(For protocol 2 the variation of structures is somewhat larger resulting in an average RMSD of 1.02. If we combine the $1,000$ runs from protocol $1$ and $2$ the average RMSD increases slightly to $1.172$ indicating
that structures from protocol $1$ and $2$ are not identical but very similar.)
More formally one can also construct a trihedron from the center of masses of the overall structure and the first three chromosomes. The (somewhat arbitrarily defined) chirality can then be determined, for example, 
by computing the sign of the corresponding vector product.
\begin{figure}[h]
	\centering
	\includegraphics[width=0.4\textwidth]{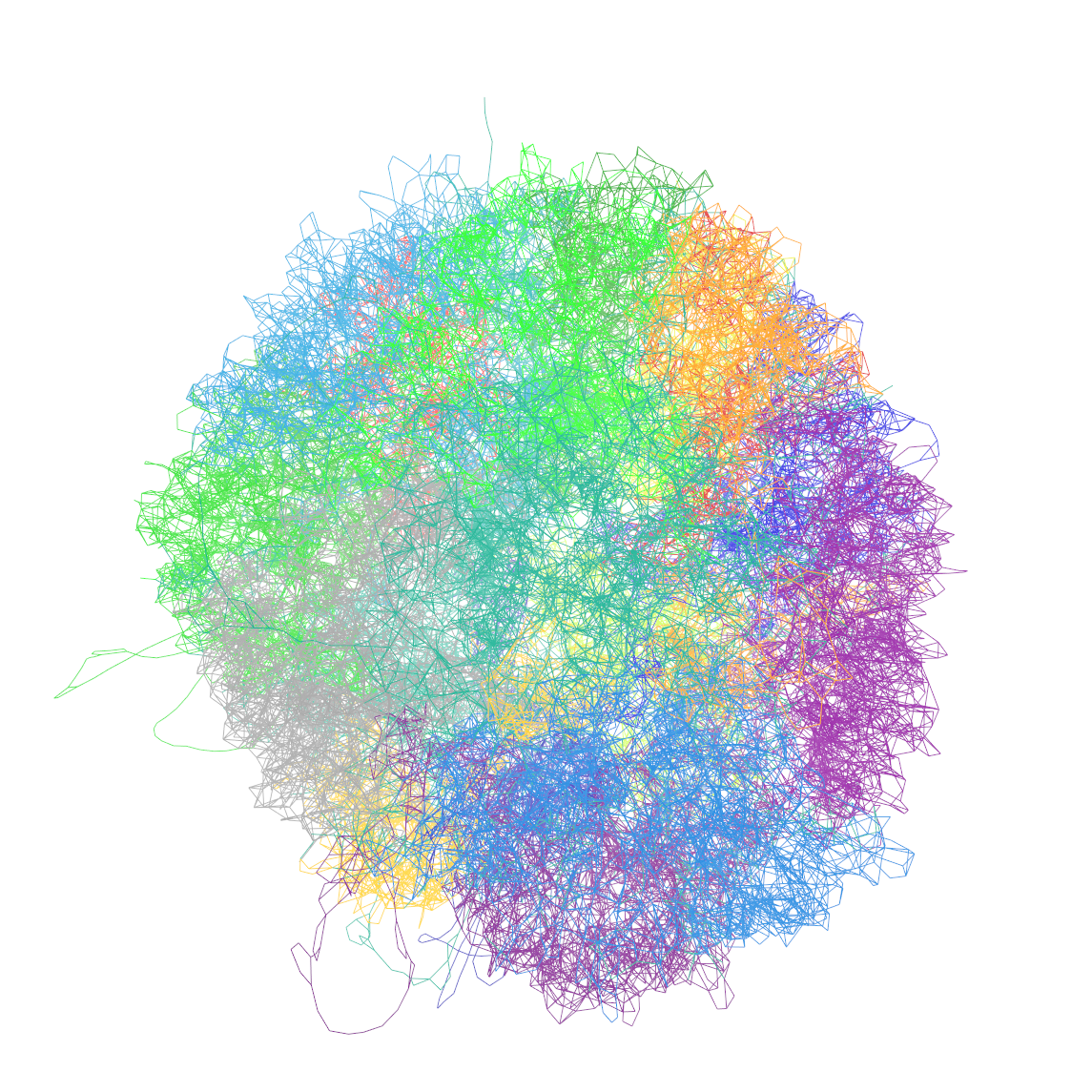}
	\includegraphics[width=0.4\textwidth]{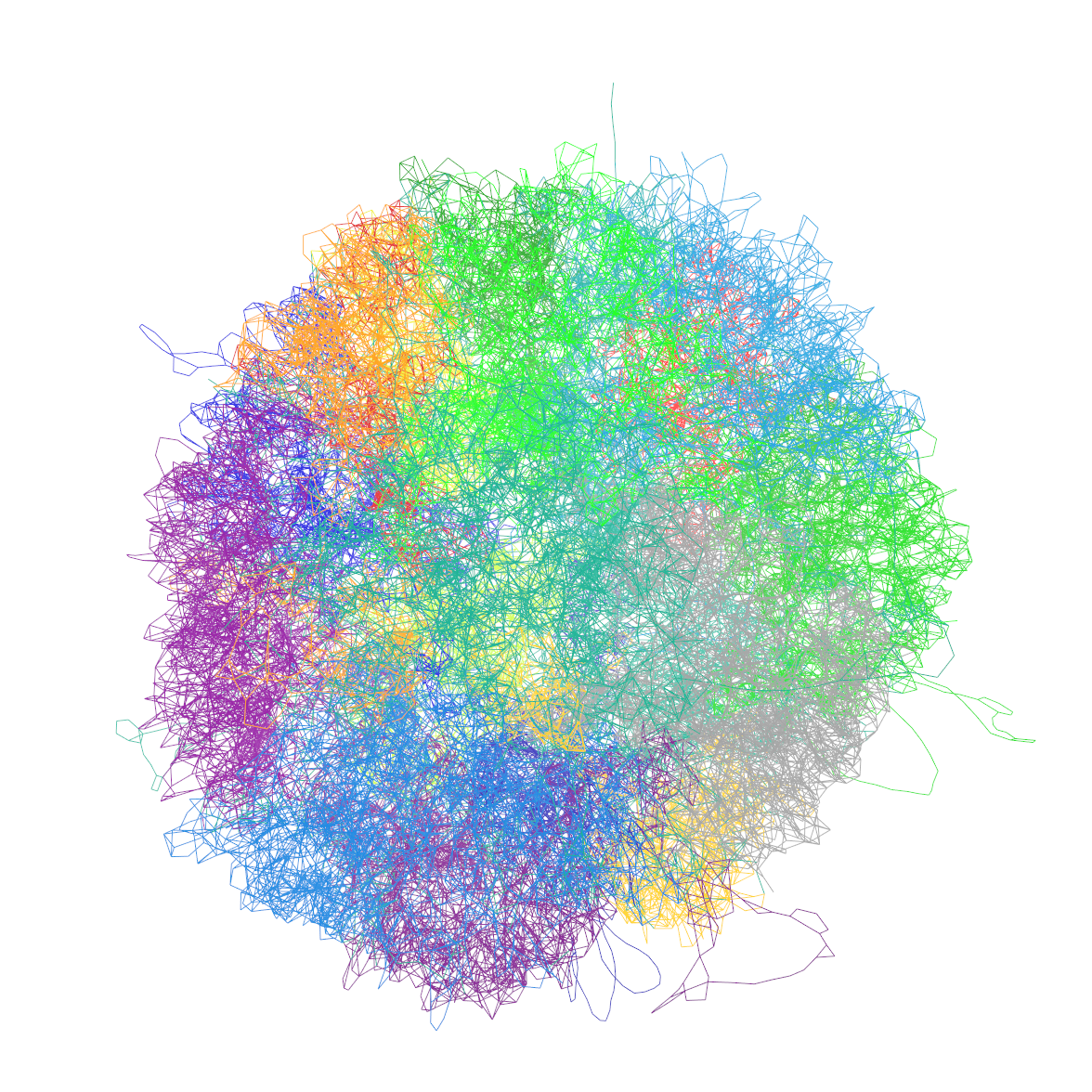}
	\caption{Examples of two chiral structures emerging from protocol 2. Different colors represent the individual chromosomes. 
	Figures were generated with VMD \cite{VMDPaper:1996}.}
        \label{fig:chiral_confs}
\end{figure}
To summarize, our empirical observation indicates that information from contact maps is insufficient to determine the chirality of the underlying structure which reflects the inherent symmetry of the associated matrix.
The emergence of mirror images in the modeling of interphase chromatin has already been described in \cite{CarstensPLoS:2016} for single chromosomes, but to our knowledge not for whole genome structures. 
Note that the true chirality of the cell may always be obtained by other means, for example, if a microscope image of the same cell exists.

\section{Knots}

It is generally assumed that interphase chromosomes are (mostly) unknotted. It has been argued that knots may lead to breakage of DNA \cite{Deibler_BMC:2007} or hinder 
RNA transcription \cite{PortugalNAR:1996}, and recently, mechanisms have been suggested to actively unknot chromosomes via type II DNA topoisomerases \cite{RackoPolymers:2018}. 
The theoretical foundation for the presumed absence of knots is provided by the so-called crumpled globule model, a concept developed in the $1980'$s \cite{GrosbergJP:1988, GrosbergEPL:1993} in the context of polymer physics. 
Essentially, interphase chromosomes are believed to be akin to an unknotted polymer which after a collapse from an extended conformation remains unknotted in a non-equilibrium metastable state. The main argument for
the latter is the power-law decay of the contact probability which exhibits a similar exponent as a crumpled globule \cite{LiebermannScience2009}. 
It has also been shown that chains with an extensive degree of knottedness can be constructed while still preserving the correct exponent \cite{ArsuagaFront2015}. 

Recently, knots have been observed directly
in circular minichromosomes of yeast \cite{ValdesNAR:2018}, and we have analyzed three-dimensional structures of haploid interphase chromosomes of mice based on the same data and a similar model as 
the one discussed in this paper. This analysis revealed that these
structures are knotted to a substantial degree \cite{Siebert:2017}. However, different optimization runs based on data from the same cell resulted in similar, but somewhat different topologies. Even though evidence was 
presented that at least some of these knots are likely real and not artifacts of the modeling, the question whether or not knots exist in interphase chromosomes could not be resolved conclusively. 
Other modeling attempts based on data from multiple cells have also detected knots, however, to a somewhat lesser degree \cite{PierroPNAS:2016,ZhangPNAS:2015}. In the following paragraph we would like to shed some more light on this 
controversial issue.

To gauge the emergence of knots we will focus on chromosome $14$ from cell $2$ which has already been discussed in great detail in our previous publication \cite{Siebert:2017}. Table~\ref{knot_prob} displays the
occurrence of knots in the structures constructed with protocol $1$ and $2$ in comparison to our reference model \cite{StevensNature:2017}. The latter is based on the same Hi-C data but uses somewhat different 
potentials and minimization protocols. Knots are determined by closing the ends via an extension of two lines through the center of mass and the termini of the chromosome before computing a variant of the
Alexander polynomial as described in detail in Ref.~\cite{Siebert:2017}. Note that details of the closure may have a minor influence on the probabilities \cite{VirnauJACS:2006, VirnauPLOS:2006} of the observed knots. We have opted for a rather simple
closure which has been used successfully in the context of proteins \cite{VirnauPLOS:2006}. However, we have also checked that results do not change qualitatively if more sophisticated closures such as variants of the statistical closure 
\cite{VirnauPLOS:2006, Millett:2013}
are applied in agreement with previous results \cite{VirnauJACS:2006, VirnauPLOS:2006}.
 Structures can in principle also be analyzed using a dedicated web server which was set up to check models of chromatin 
for knots \cite{Sulkowska_NAR:2018}.

\begin{table}
 \centering
    \begingroup
    \renewcommand{\arraystretch}{1.5}
    \begin{tabular}{|l|l|l|l|}
     \hline
	    & \textbf{protocol 1} & \textbf{protocol 2} & \textbf{Ref. \cite{StevensNature:2017}} \\
     \hline
     \text{number of configurations} & 500 & 500 & 10 \\
     \hline
     0 & 0 & 0.002 & 0 \\
     \hline
     $3_{1}$ & 0.328 & 0.342 & 0.4 \\
     \hline
     $4_{1}$ & 0 & 0 & 0 \\
     \hline
     $5_{1}$ & 0 & 0 & 0 \\
     \hline
     $5_{2}$ & 0 & 0 & 0 \\
     \hline
     $\#3_{1} \# 3_{1}$ & 0.31 & 0.21 & 0.3 \\
     \hline
     other & 0.362 & 0.446 & 0.3 \\
     \hline
     \hline
     fraction of $3_{1}$ at [290,340] & 1.0 & 0.918 & 1.0 \\
     \hline
    \end{tabular}
\endgroup
\vspace{0.5cm}
	\caption{Knotting probabilities for chromosome $14$ of cell $2$ for the unknot (0), the trefoil (3$_1$), the figure-eight knot (4$_1$), the two five-fold knots and a composit knot made up of two trefoils. 
	The last row indicates the fraction of trefoil knots located in between bead position 290 and 340.}
	\label{knot_prob}
\end{table}

As in the reference model we are not able to steer the minimization towards a unique topological state but observe a variety of simple knots. Note however, that this task is far from trivial as a small
displacement of a single bond suffices to change knot types as demonstrated in \cite{Siebert:2017}. Nevertheless, we observe very similar knotting probabilities and a noteworthy absence of unknots, 
figure-eight and five-fold knots in all cases. 
Surprisingly, the simple trefoil knot located between monomer number 290 and 340 (Fig.~\ref{fig:chiral_confs}) is present in (almost) all structures (last row in Table~\ref{knot_prob}), 
irrespective of the simulation protocol and even the underlying model.
This not only provides further evidence that this particular knot actually existed in the cell on which the Hi-C map is based upon, but also for the consistency of our approach. We have also checked that this particular knot is
not present in any other cell lines published in \cite{StevensNature:2017}. Together with the seemingly different structures emerging from the different cells in \cite{StevensNature:2017}, this may serve as an indication
that single-cell data is actually required to obtain reliable structural representations.

\begin{figure}[h]
	\centering
	\includegraphics[width=0.4\textwidth]{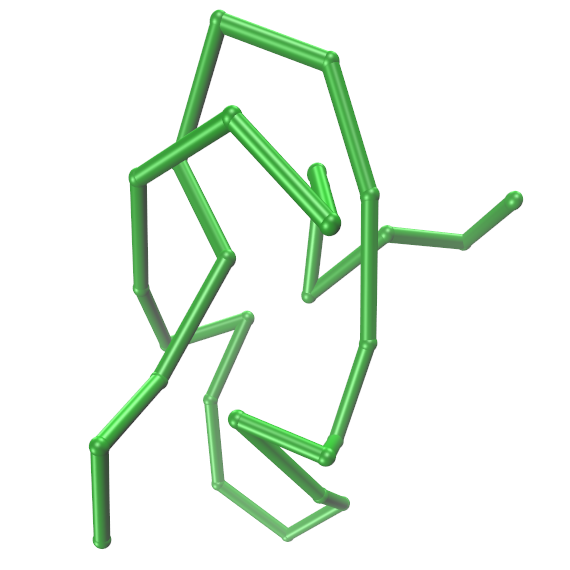}
	\includegraphics[width=0.4\textwidth]{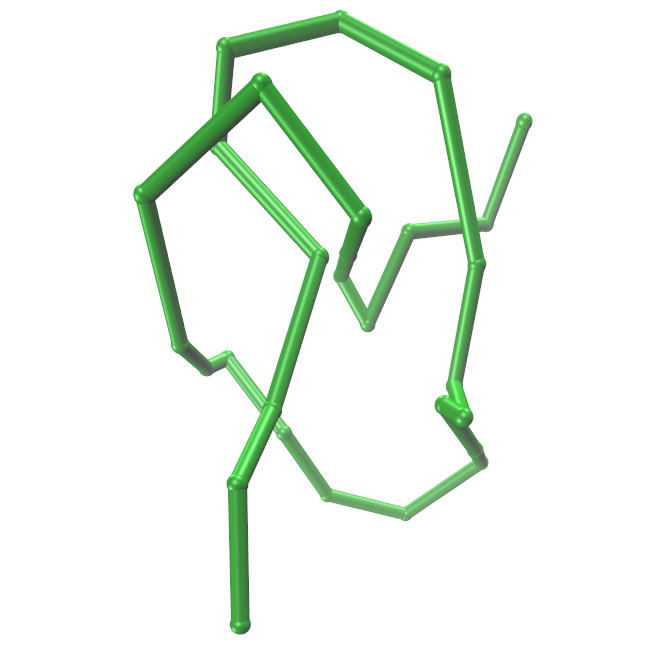}
	\caption{a$)$ Trefoil knot observed in our simulations (cell 2, chromsome 14, monomer number 301 to 329, protocol 1). b$)$ Same trefoil observed in structure of Ref.~\cite{StevensNature:2017} (model 5).}
        \label{fig:chiral_confs}
\end{figure}

Finally, we would like to point out an observation from our simulations. The amount and complexity of knots strongly depends on the excluded volume contribution of the potential. States with no or little excluded volume
as produced by steps 1 and 2 of protocol 1 are knotted beyond the capabilities of our algorithm to distinguish them or in fact any algorithm we are aware of. The additional constraints due to contacts increase knotting 
substantially beyond ideal chains or equilibrium globules \cite{VirnauJACS:2006}. Only a strong excluded volume term guarantees the modest knotting observed in Table~\ref{knot_prob}. From this point of view knotting does depend 
on the details of the potentials and Table~\ref{knot_prob} represents in a way the lowest degree of knotting we were able to obtain with our ansatz. Also note that an essentially unknotted starting configuration 
(such as created from a random walk in the first few time steps of step 1 of protocol 2) does not guarantee that final configurations are unknotted as long as bond crossings can occur.

\section{Summary and Outlook}
In this paper we present a minimal structure-based model, which in conjunction with a highly efficient minimization protocol, allows for the instantaneous determination of three-dimensional structures of whole genomes from
haploid Hi-C contact maps at the level of 100,000 base pairs per monomer. Irrespective of protocol details and starting conditions, structures appear to converge to conformationally similar low energy states as long as bond 
crossings are allowed. We find evidence that 
the mere fulfillment of contact maps is insufficient to create experimentally meaningful structures and that an excluded volume potential which essentially enforces non-contacts is required for the formation of chromosome 
territories. Furthermore, contact maps also do not contain information about the chirality of the underlying structures, so that one of two mirror images may emerge from our minimization runs. Finally, we also investigate 
self-entanglements in a particular chromosome. We observe that the same trefoil knot is present in (almost) all of our structures as well as in structures which employ a completely different optimization strategy and potentials, 
providing further evidence for the consistency of our approach and for the existence of knots in interphase chromatin.

In future we would like to extend our modeling ansatz to higher resolutions while ensuring consistency across all levels of coarse-graining.
We also intend to apply our approach to the conceptually challenging modeling of diploid chromosomes \cite{TanScience:2018}.

\section*{Acknowledgements}
This project is funded by the Deutsche Forschungsgemeinschaft (DFG, German Research Foundation) - project number 233630050 - TRR 146.

\clearpage

\bibliographystyle{ieeetr}
\bibliography{bibliography}

\begin{thebibliography}{10}

\bibitem{CremerNRG:2001}
T.~Cremer and C.~Cremer, ``Chromosome territories, nuclear architecture and
  gene regulation in mammalian cells,'' {\em {N}ature {R}eviews {G}enetics},
  no.~4, pp.~292--301, 2001.

\bibitem{DixonNature:2015}
J.~R. Dixon, I.~Jung, S.~Selvaraj, Y.~Shen, J.~E. Antosiewicz-Bourget, A.~Y.
  Lee, Z.~Ye, A.~Kim, N.~Rajagopal, W.~Xie, Y.~Diao, J.~Liang, H.~Zhao, V.~V.
  Lobanenkov, J.~R. Ecker, J.~A. Thomson, and B.~Ren, ``Chromatin architecture
  reorganization during stem cell differentiation,'' {\em Nature}, vol.~518,
  pp.~331--336, 2015.

\bibitem{NaumovaScience:2013}
N.~Naumova, M.~Imakaev, G.~Fudenberg, Y.~Zhan, B.~R. Lajoie, L.~A. Mirny, and
  J.~Dekker, ``Organization of the mitotic chromosome,'' {\em Science},
  vol.~342, no.~6161, pp.~948--953, 2013.

\bibitem{CremerCSH:1993}
T.~Cremer, A.~Kurz, R.~M. Zirbel, S.~Dietzel, B.~Rinke, E.~Schr\"ock,
  M.~Speicher, U.~Mathieu, A.~Jauch, P.~Emmerich, H.~Scherthan, T.~Ried,
  C.~Cremer, and P.~Lichter, ``Role of chromosome territories in the functional
  compartmentalization of the cell nucleus,'' {\em Cold Spring Harbor Symposia
  on Quantitative Biology}, vol.~58, pp.~777--92, 02 1993.

\bibitem{BrancoPLoS:2006}
M.~R. Branco and A.~Pombo, ``Intermingling of chromosome territories in
  interphase suggests role in translocations and transcription-dependent
  associations,'' {\em PLOS Biology}, vol.~4, no.~5, pp.~780--788, 2006.

\bibitem{DekkerScience:2009}
J.~Dekker, K.~Rippe, M.~Dekker, and N.~Kleckner, ``Capturing chromosome
  conformation,'' {\em Science}, vol.~295, no.~5558, pp.~1306--1311, 2002.

\bibitem{LiebermannScience2009}
E.~Lieberman-Aiden, N.~L. van Berkum, L.~Williams, M.~Imakaev, T.~Ragoczy,
  A.~Telling, I.~Amit, B.~R. Lajoie, P.~J. Sabo, M.~O. Dorschner, R.~Sandstrom,
  B.~Bernstein, M.~Bender, M.~Groudine, A.~Gnirke, J.~Stamatoyannopoulos, L.~A.
  Mirny, E.~S. Lander, and J.~Dekker, ``Comprehensive mapping of long-range
  interactions reveals folding principles of the human genome,'' {\em Science},
  vol.~326, no.~5950, pp.~289--293, 2009.

\bibitem{NaganoNature:2013}
T.~Nagano, Y.~Lubling, T.~Stevens, S.~Schoenfelder, E.~Yaffe, W.~Dean, E.~Laue,
  A.~Tanay, and P.~Fraser, ``Single-cell {Hi}-{C} reveals cell-to-cell
  variability in chromosome structure,'' {\em Nature}, vol.~502, pp.~59--64, 09
  2013.

\bibitem{StevensNature:2017}
T.~J. Stevens, D.~Lando, S.~Basu, L.~P. Atkinson, Y.~Cao, S.~F. Lee, M.~Leeb,
  K.~J. Wohlfahrt, W.~Boucher, A.~O’Shaughnessy-Kirwan, J.~Cramard, A.~J.
  Faure, M.~Ralser, E.~Blanco, L.~Morey, M.~Sans{\'o}, M.~G.~S. Palayret,
  B.~Lehner, L.~D. Croce, A.~Wutz, B.~Hendrich, D.~Klenerman, and E.~D. Laue,
  ``3{D} structures of individual mammalian genomes studied by single-cell
  {H}i-{C},'' {\em Nature}, vol.~544, pp.~59--64, 2017.

\bibitem{NaganoNature:2017}
T.~Nagano, Y.~Lubling, C.~V{\'a}rnai, C.~Dudley, W.~Leung, Y.~Baran, N.~M.
  Cohen, S.~Wingett, P.~Fraser, and A.~Tanay, ``Cell-cycle dynamics of
  chromosomal organization at single-cell resolution,'' {\em Nature}, vol.~547,
  pp.~61--67, 2017.

\bibitem{FlyamerNature:2017}
I.~M. Flyamer, J.~Gassler, M.~Imakaev, H.~B. Brand{\~a}o, S.~V. Ulianov,
  N.~Abdennur, R.~S.~V, L.~A. Mirny, and K.~Tachibana-Konwalski,
  ``Single-nucleus {H}i-{C} reveals unique chromatin reorganization at
  oocyte-to-zygote transition,'' {\em Nature}, vol.~544, pp.~110 --114, 2017.

\bibitem{RamaniNM:2017}
V.~Ramani, X.~Deng, R.~Qiu, K.~L. Gunderson, F.~J. Steemers, C.~M. Disteche,
  W.~S. Noble, Z.~Duan, and J.~Shendure, ``Massively multiplex single-cell
  {H}i-{C},'' {\em Nature Methods}, vol.~14, pp.~263 --366, 2017.

\bibitem{TanScience:2018}
L.~Tan, D.~Xing, C.-H. Chang, H.~Li, and X.~S. Xie, ``Three-dimensional genome
  structures of single diploid human cells,'' {\em Science}, vol.~361,
  no.~6405, pp.~924--928, 2018.

\bibitem{Rosenthal:2018}
M.~Rosenthal, D.~Bryner, F.~Huffer, S.~Evans, A.~Srivastava, and N.~Neretti,
  ``Bayesian estimation of 3{D} chromosomal structure from single cell {H}i-{C}
  data,'' {\em bioRxiv}, 2018.

\bibitem{PaulsenPLoS:2015}
J.~Paulsen, O.~Gramstad, and P.~Collas, ``Manifold based optimization for
  single-cell 3{D} genome reconstruction,'' {\em PLOS Computational Biology},
  vol.~11, no.~8, pp.~1--19, 2015.

\bibitem{CarstensPLoS:2016}
S.~Carstens, M.~Nilges, and M.~Habeck, ``Inferential structure determination of
  chromosomes from single-cell {H}i-{C} data,'' {\em PLOS Computational
  Biology}, vol.~12, no.~12, pp.~1--33, 2016.

\bibitem{ZhangPNAS:2015}
B.~Zhang and P.~G. Wolynes, ``Topology, structures, and energy landscapes of
  human chromosomes,'' {\em Proceedings of the National Academy of Sciences},
  vol.~112, no.~19, pp.~6062--6067, 2015.

\bibitem{ZhangPRL:2016}
B.~Zhang and P.~G. Wolynes, ``Shape transitions and chiral symmetry breaking in
  the energy landscape of the mitotic chromosome,'' {\em Phys. Rev. Lett.},
  vol.~116, p.~248101, 2016.

\bibitem{ZhangBiophys:2017}
B.~Zhang and P.~G. Wolynes, ``Genomic energy landscapes,'' {\em Biophysical
  Journal}, vol.~112, no.~3, pp.~427--433, 2017.

\bibitem{StefanoPLoS:2013}
M.~D. Stefano, A.~Rosa, V.~Belcastro, D.~di~Bernardo, and C.~Micheletti,
  ``Colocalization of coregulated genes: A steered molecular dynamics study of
  human chromosome 19,'' {\em PLOS Computational Biology}, vol.~9, no.~3,
  pp.~1--13, 2013.

\bibitem{StefanoSR:2016}
M.~D. Stefano, J.~Paulsen, T.~G. Lien, E.~Hovig, and C.~Micheletti,
  ``Hi-{C}-constrained physical models of human chromosomes recover
  functionally-related properties of genome organization,'' {\em Scientific
  Reports}, vol.~6, p.~35985, 2016.

\bibitem{PierroPNAS:2016}
M.~D. Pierro, B.~Zhang, E.~L. Aiden, P.~G. Wolynes, and J.~N. Onuchic,
  ``Transferable model for chromosome architecture,'' {\em Proceedings of the
  National Academy of Sciences}, vol.~113, no.~43, pp.~12168--12173, 2016.

\bibitem{Taketomi:1975}
H.~Taketomi, Y.~Ueda, and N.~G\=o, ``Studies on protein folding, unfolding and
  fluctuations by computer simulation,'' {\em International Journal of Peptide
  and Protein Research}, vol.~7, no.~6, pp.~445--459, 1975.

\bibitem{Socci:1995}
J.~D. Bryngelson, J.~Onuchic, N.~Socci, and P.~Wolynes, ``Funnels, pathways,
  and the energy landscape of protein folding: A synthesis,'' {\em Proteins},
  vol.~21, no.~3, pp.~167--195, 1995.

\bibitem{Boelinger:2010}
D.~B\"olinger, J.~I. Sulkowska, H.-P., L.~A. Mirny, M.~Kardar, J.~Onuchic, and
  P.~Virnau, ``A {S}tevedore's protein knot,'' {\em PLOS Computational
  Biology}, vol.~6, no.~4, pp.~1--6, 2010.

\bibitem{JarmolinskaJMB:2019}
A.~I. Jarmolinska, A.~P. Perlinska, R.~R. B., Trefz, H.~M. Ginn, P.~Virnau, and
  J.~I. Sulkowska, ``Proteins' knotty problems,'' {\em Journal of Molecular
  Biology}, vol.~431, pp.~244--257, 2019.

\bibitem{ProteinDataBank:2000}
H.~M. Berman, J.~Westbrook, Z.~Feng, G.~Gilliland, T.~N. Bhat, H.~Weissig,
  I.~N. Shindyalov, and P.~E. Bourne, ``The protein data bank,'' {\em Nucleic
  Acids Research}, vol.~28, no.~1, pp.~235--242, 2000.

\bibitem{Auhl:2003}
R.~Auhl, R.~Everaers, G.~S. Grest, K.~Kremer, and S.~J. Plimpton,
  ``Equilibration of long chain polymer melts in computer simulations,'' {\em
  The Journal of Chemical Physics}, vol.~119, no.~24, pp.~12718--12728, 2003.

\bibitem{Kirkpatrick671}
S.~Kirkpatrick, C.~D. Gelatt, and M.~P. Vecchi, ``Optimization by simulated
  annealing,'' {\em Science}, vol.~220, no.~4598, pp.~671--680, 1983.

\bibitem{KremerGrest:1990}
K.~Kremer and G.~S. Grest, ``Dynamics of entangled linear polymer melts: A
  molecular-dynamics simulation,'' {\em The Journal of Chemical Physics},
  vol.~92, no.~8, pp.~5057--5086, 2017.

\bibitem{Diao:2015}
Y.~Diao, C.~Ernst, S.~Saarinen, and U.~Ziegler, ``Generating random walks and
  polygons with stiffness in confinement,'' {\em Journal of Physics A:
  Mathematical and Theoretical}, vol.~48, p.~095202, 2015.

\bibitem{NilgesNMR:1988}
M.~Nilges, G.~M. Clore, and A.~M. Gronenborn, ``Determination of
  three-dimensional structures of proteins from interproton distance data by
  dynamical simulated annealing from a random array of atoms circumventing
  problems associated with folding,'' {\em FEBS Letters}, vol.~239, no.~1,
  pp.~129--136, 1988.

\bibitem{VMDPaper:1996}
W.~Humphrey, A.~Dalke, and K.~Schulten, ``{VMD} -- {V}isual {M}olecular
  {D}ynamics,'' {\em Journal of Molecular Graphics}, vol.~14, pp.~33--38, 1996.

\bibitem{Anderson:2008}
J.~A. Anderson, C.~D. Lorenz, and A.~Travesset, ``General purpose molecular
  dynamics simulations fully implemented on graphics processing units,'' {\em
  Journal of Computational Physics}, vol.~227, no.~10, pp.~5342--5359, 2008.

\bibitem{Glotzer:2015}
J.~Glaser, T.~D. Nguyen, J.~A. Anderson, P.~Lui, F.~Spiga, J.~A. Millan, D.~C.
  Morse, and S.~C. Glotzer, ``Strong scaling of general-purpose molecular
  dynamics simulations on {GPU}s,'' {\em Computer Physics Communications},
  vol.~192, pp.~97--107, 2015.

\bibitem{LBVH:2016}
M.~P. Howard, J.~A. Anderson, A.~Nikoubashman, S.~C. Glotzer, and A.~Z.
  Panagiotopoulos, ``Efficient neighbor list calculation for molecular
  simulation of colloidal systems using graphics processing units,'' {\em
  Computer Physics Communications}, vol.~203, pp.~45--52, 2016.

\bibitem{Deibler_BMC:2007}
R.~W. Deibler, J.~K. Mann, W.~L. de~Sumners, and L.~Zechiedrich, ``Hin-mediated
  {DNA} knotting and recombining promote replicon dysfunction and mutation.,''
  {\em BMC Molecular Biology}, vol.~8, no.~1, p.~44, 2007.

\bibitem{PortugalNAR:1996}
J.~Portugal and A.~Rodríguez-Campos, ``T7 {RNA} polymerase cannot transcribe
  through a highly knotted {DNA} template,'' {\em Nucleic Acids Research},
  vol.~24, no.~24, pp.~4890--4894, 1996.

\bibitem{RackoPolymers:2018}
D.~Racko, F.~Benedetti, D.~Goundaroulis, and A.~Stasiak, ``Chromatin-loop
  extrusion and chromatin unknotting,'' {\em Polymers}, vol.~10, 2018.

\bibitem{GrosbergJP:1988}
A.~Y. Grosberg, S.~Nechaev, and E.~I. Shakhnovich, ``The role of topological
  constraints in the kinetics of collapse of macromolecules,'' {\em J. Phys.
  France}, vol.~49, no.~12, pp.~2095--2100, 1988.

\bibitem{GrosbergEPL:1993}
A.~Y. Grosberg, Y.~Rabin, S.~Havlin, and A.~Neer, ``Crumpled globule model of
  the three-dimensional structure of {DNA},'' {\em Europhysics Letters
  ({EPL})}, vol.~23, no.~5, pp.~373--378, 1993.

\bibitem{ArsuagaFront2015}
J.~Arsuaga, R.~Jayasinghe, R.~Scharein, M.~Segal, R.~Stolz, and M.~Vazquez,
  ``Current theoretical models fail to predict the topological complexity of
  the human genome,'' {\em Frontiers in Molecular Biosciences}, vol.~2, p.~48,
  2015.

\bibitem{ValdesNAR:2018}
A.~Vald{\'e}s, J.~Segura, S.~Dyson, B.~Mart{\'i}nez-Garc{\'i}a, and J.~Roca,
  ``{DNA} knots occur in intracellular chromatin,'' {\em Nucleic Acids
  Research}, vol.~46, no.~2, pp.~650--660, 2017.

\bibitem{Siebert:2017}
J.~T. Siebert, A.~N. Kivel, L.~P. Atkinson, T.~Stevens, E.~Laue, and P.~Virnau,
  ``Are there knots in chromosomes?,'' {\em Polymers}, vol.~9, p.~317, 2017.

\bibitem{VirnauJACS:2006}
P.~Virnau, Y.~Kantor, and M.~Kardar, ``Knots in globule and coil phases of a
  model polyethylene,'' {\em Journal of the American Chemical Society},
  vol.~127, pp.~15102--6, 2005.

\bibitem{VirnauPLOS:2006}
P.~Virnau, L.~A. Mirny, and M.~Kardar, ``Intricate knots in proteins: functions
  and evolution,'' {\em PLOS Computational Biology}, vol.~2, no.~9,
  pp.~1074--1079, 2006.

\bibitem{Millett:2013}
K.~C. Millett, E.~J. Rawdon, A.~Stasiak, and J.~I. Sulkowska, ``Identifying
  knots in proteins,'' {\em Biochemical Society Transactions}, vol.~41, no.~2,
  pp.~533--537, 2013.

\bibitem{Sulkowska_NAR:2018}
J.~I.Sulkowska, S.~Niewieczerzal, A.~I. Jarmolinska, J.~T. Siebert, P.~Virnau,
  and W.~Niemyska, ``Knotgenome: a server to analyze entanglements of
  chromosomes,'' in {\em Nucleic Acids Research}, pp.~W17--W24, 2018.

\end{thebibliography}

\end{document}